\documentclass[a4paper,pra,aps,amssymb,amsmath,showpacs,reprint]{revtex4-1}
\usepackage{graphicx}
\usepackage{bm}
\usepackage{amsmath}
\usepackage{dcolumn}
\usepackage{color}
\usepackage{hyperref}
\usepackage{lipsum}  

\hypersetup{
   pdfpagemode=UseNone,
   pdfstartpage=1,
   pdfstartview=FitH,
   pdfmenubar=true,
   pdftoolbar=true,
   colorlinks = true,
   linkcolor=blue,
   citecolor=blue,
   bookmarksopen=false
}

\newcolumntype{d}[1]{D{.}{.}{#1}}
\newcolumntype{e}[1]{D{.}{\;\;}{#1}}

\newcommand{\dd}{\mathrm{d}}

\newcommand{\hartree}{E$_\mathrm{h}$}

\begin{document}

\title{Correlation energies for many-electron atoms with explicitly correlated Slater functions}

\author{Micha{\l} Przybytek}
\email{michal.przybytek@tiger.chem.uw.edu.pl}

\author{Micha{\l} Lesiuk}
\affiliation{Faculty of Chemistry, University of Warsaw, Pasteura 1, 02-093 Warsaw, Poland}

\date{\today}

\begin{abstract}
In this work we propose a novel composite method for accurate calculation of the energies of many-electron atoms. The
dominant contribution to the energy (pair energies) are calculated by using explicitly correlated factorisable coupled
cluster theory. Instead of the usual Gaussian-type geminals for the expansion of the pair functions, we employ
two-electron Hylleraas basis set. This eliminates the need for massive optimisation of nonlinear parameters and the
required three-electron integrals can now be calculated relatively easily. The remaining contributions to the energy
are calculated within the algebraic approximation by using large one-electron basis sets composed of Slater-type
orbitals. The method is tested for the beryllium atom where the accuracy better than $1\,$cm$^{-1}$ is obtained. We
discuss in details possible sources of the error and estimate the uncertainty in each energy component. Finally, we
consider possible strategies to improve the accuracy of the method by one to two orders of magnitude. The most
important advantage of the method is that is does not suffer from an exponential growth of the computational costs with
increasing number of electrons in the system and thus can be applied to heavier atoms preserving a similar level of
accuracy.
\end{abstract}

\maketitle

\section{Introduction}

Atomic spectroscopy remains an important and active field of modern physics. Many theoretical and experimental works
concentrated on different aspects of atomic spectra touch upon very fundamentals of the present scientific knowledge.
Search for time-reversal symmetry violations \cite{roberts15}, time-dependence of fundamental physical constants \cite{rosenband08,hunt14,godun14}, various empirical tests of the Standard Model and quantum electrodynamics \cite{eikema97,korobov01,pachucki06,odom06,gabrielse07} are only a handful of prominent examples. Therefore, the need for
development of new accurate theoretical tools to predict the atomic spectra (and other relevant quantities) is easy to
recognise.

If we restrict ourselves to light atoms, the most accurate theoretical results to date have been obtained with
methods where all inter-particle distances are explicitly incorporated into the trial wavefunction. This includes basis
sets of Hylleraas-type functions \cite{hylleraas29a,hylleraas29b,drake94,puchalski06}, explicitly-correlated Gaussians (ECG) \cite{szalewicz10,mitroy13}, Hylleraas-CI expansions \cite{sims71,sims07,king93}, and Slater geminals \cite{thakkar77,frolov95, korobov02,puchalski10}. The common
problem among these methods, however, is the exponential scaling of the computational costs with the number of particles
in the system. Applications to systems larger than, say, five particles are thus scarce and much less accurate.

A different approach to the electronic structure is offered by the coupled clusters (CC) theory \cite{bartlett07}. Since the total CC
wavefunction is parametrised in terms of a cluster operator which can be truncated in a systematic way, the exponential
increase of the computational costs is avoided. The most popular implementation of the CC theory relies on the algebraic
approximation, i.e. expansion in a set of one-electron orbitals. Unfortunately, this leads to a relatively slow
convergence of the results towards the complete basis set limit \cite{hill85} - a manifestation of the Kato's electron-electron cusp condition \cite{kato57}.

One possible remedy to this problem is to abandon the algebraic approximation entirely. To this end, various authors
showed that a basis-set independent CC theory can be formulated in terms of the so-called pair functions \cite{byron66,pan70,pan72,chalas77,szalewicz79,szalewicz82,szalewicz83a,szalewicz83b,szalewicz84a,szalewicz84b,adamowicz77,adamowicz78a,adamowicz78b}. The pair
functions are two-electron objects and thus can be expanded in a basis set which overtly includes all coordinates of
the given electron pair. This idea gave rise to the explicitly correlated CC theory. 

One of the most difficult obstacles preventing straightforward application of this method is the presence of many-electron integrals. In the modern R12/F12 theory \cite{hattig12,kong12,tenno12} this difficulty is
avoided by proper insertions of the resolution of identity (RI) approximation. A different idea has been proposed by
Szalewicz and collaborators \cite{szalewicz82,szalewicz83a,szalewicz83b} who imposed the strong-orthogonality requirement for the pair functions only in the complete basis set limit.  This led to a
family of weak-orthogonality (WO) functionals. At the second-order M\o ller-Plesset (MP2) level of theory \cite{bartlett07}, for example, this eliminates all four-electron integrals from the working equations, leaving only the relatively simple three-electron ones \cite{szalewicz82,szalewicz83a}.

The WO functionals are typically combined with the Gaussian-type geminals (GTG) for expansion of the
CC pair functions. The main advantage of such approach is that the resulting three-electron integrals can be evaluated
analytically in a closed form. The remaining inconvenience of the WO theory is the need for
optimisation of GTG nonlinear parameters. Despite a considerable effort it seems that
this problem has not been satisfactorily resolved thus far. Massive nonlinear optimisations can take months or even
years of computer time to converge to a satisfactory result.

With that said, for one-centre systems one can entertain an idea of using the Hylleraas basis set of the form 
\begin{equation}
\label{eq:postac}
(4\pi)^{-1}r_1^{u}\,r_2^{v}\,r_{12}^{t}\,\exp(-a_i\,r_1-a_j\,r_2),
\end{equation}
where $a_i,a_j>0$, and $u$, $v$, $t$ are non-negative integers, for the expansion of the CC pair functions. 
As explained further in the text, the most important advantage of this basis set is that instead of hundreds or thousands of nonlinear parameters per electron pair one has only a handful of them. Moreover, the basis set (\ref{eq:postac}) can be systematically extended so that the basis set limits and the corresponding errors bars are easier to estimate.

The basis set (\ref{eq:postac}) has not found a significant use in CC theory thus far because of the resulting
three-electron integrals. However, in the past two decades a considerable progress has been achieved in attempts to
evaluate them analytically and/or recursively. This started with the seminal paper of Fromm and Hill \cite{fromm87} who solved the
simplest three-electron integral with inverse powers of all interparticle distances. Despite this success and
subsequent works \cite{remiddi91,harris97} the analytic formulae were lengthy and their evaluation (and differentiation) both expensive and prone
to numerical instabilities. 

Somewhat later, Pachucki and collaborators \cite{pachucki04a} proposed a set of recursive formulae connecting all three-electron integrals
resulting from the basis set (\ref{eq:postac}), thereby eliminating many problems shared by the previous approaches.
This opens up a new avenue for application of the WO CC theory to many-electron atoms. Since the three-electron
integrals are no longer a bottleneck, the basis set (\ref{eq:postac}) is expected to be superior to the GTG expansion,
both in terms of accuracy (satisfies the cusp condition) and computational efficiency (a small number of independent
nonlinear parameters).

\section{Theory and implementation}

\subsection{Explicitly correlated calculations}

In the first-quantised formulation of the factorisable coupled cluster doubles theory (FCCD) \cite{szalewicz84a} the electron correlation effects in an
$N$-electron closed-shell system 
are expressed in terms of a set of $N^2/4$ spinless pair functions of well-defined permutational symmetry.
There are $[N(N/2+1)]/4$ independent singlet pair functions $\tau^1_{\alpha\beta}(1,2)$ 
which are symmetric with respect to the exchange of electron coordinates
and orbital indices $\alpha$ and $\beta$, and $[N(N/2-1)]/4$ triplet pair functions $\tau^3_{\alpha\beta}(1,2)$ which
are antisymmetric under these operations,
i.e. $\tau^s_{\alpha\beta}(1,2)=(2-s)\tau^s_{\alpha\beta}(2,1)=(2-s)\tau^s_{\beta\alpha}(1,2)$, $s=1,3$.

We assume that the reference Hartree-Fock determinant is constructed from canonical orbitals $\phi_\alpha$,
$\alpha=1\dots N/2$
(corresponding to the lowest orbital energies $\epsilon_\alpha$) which are eigenfunctions of the standard closed-shell
Fock operator $f$, i.e. $f \phi_\alpha = \epsilon_\alpha \phi_\alpha$.
In this case, the individual pair functions $\tau^s_{\alpha\beta}(1,2)$ are solutions 
to the integro-differential FCCD equations of the general form \cite{szalewicz84a,Rychlewski:book}
\begin{equation}
\label{eq:FCCD}
\left[f(1)+f(2)-\epsilon_\alpha-\epsilon_\beta\right]\tau_{\alpha\beta}^s(1,2)=R^s_{\alpha\beta}[{\bm \tau}],
\end{equation}
with an additional requirement that the pair functions must fulfil the SO condition
\begin{equation}
\label{eq:SO}
q_2(1,2)\tau_{\alpha\beta}^s(1,2)=\tau_{\alpha\beta}^s(1,2).
\end{equation}
The exact two-electron SO projector $q_2$ in Eq.~(\ref{eq:SO}) is defined as
\begin{equation}
q_2(1,2)=(1-p(1))(1-p(2)),
\end{equation}
where the action of a projector $p$ on an arbitrary function $\chi$ is expressed in terms of the occupied orbitals
$\phi_\alpha$ as
\begin{equation}
\label{eq:pchi}
p(1)\chi(1)=\sum_{\alpha=1}^{N/2}\phi_\alpha(1)\int\phi_\alpha^\star(2)\chi(2)\,\dd 2.
\end{equation}
Once the pair functions are known, the total FCCD correlation energy is computed as a sum of contributions from
individual pairs
\begin{equation}
\label{eq:etot}
E_\mathrm{corr} = 
\sum_{\alpha=1}^{N/2}\epsilon_{\alpha\alpha}^1 + 
\sum_{\alpha<\beta}^{N/2}(\epsilon_{\alpha\beta}^1+\epsilon_{\alpha\beta}^3),
\end{equation}
where the pair energies $\epsilon_{\alpha\beta}^s$ are defined by
\begin{equation}
\label{eq:epair}
\epsilon_{\alpha\beta}^s=\frac{s}{1+\delta_{\alpha\beta}}
\langle\phi_\alpha\phi_\beta|r_{12}^{-1}|\tau_{\alpha\beta}^s\rangle.
\end{equation}
The right-hand-side term $R^s_{\alpha\beta}[{\bm \tau}]$ in Eq.~(\ref{eq:FCCD}) depends explicitly on
all pair functions (indicated by the bold symbol, ${\bm \tau}$). In the FCCD theory it consists of three
contributions
\begin{equation}
\label{eq:rhs}
R_{\alpha\beta}^s[{\bm \tau}]=-q_2(1,2)r_{12}^{-1}\phi_{\alpha\beta}^s(1,2) +L_{\alpha\beta}^s[{\bm
\tau}]+F_{\alpha\beta}^s[{\bm \tau}],
\end{equation}
where $\phi_{\alpha\beta}^s$ is a properly (anti-)symmetrised product of the occupied orbitals, i.e.
$\phi^s_{\alpha\beta}(1,2)=\phi_\alpha(1)\phi_\beta(2)+(2-s)\phi_\beta(1)\phi_\alpha(2)$,
$L_{\alpha\beta}^s[{\bm \tau}]$ collects all terms which are linear in the pair functions, 
and the so-called factorisable quadratic terms are included in $F_{\alpha\beta}^s[{\bm \tau}]$.
The detailed functional form of $L_{\alpha\beta}^s[{\bm \tau}]$ and $F_{\alpha\beta}^s[{\bm \tau}]$ is found in
Ref.~\cite{szalewicz84a,Rychlewski:book}.

The coupled clusters equations are most conveniently solved iteratively. In the simplest approach 
(the straightforward iteration procedure of Ref.~\cite{szalewicz84a}), a sequence of consecutive approximations
to the pair functions, 
${\bm \tau}^{[n]}$, is generated from an equation similar to Eq.~(\ref{eq:FCCD}) but with the right-hand-side term 
calculated using the pair functions from
the previous iteration, $R^s_{\alpha\beta}[{\bm \tau}^{[n-1]}]$. The SO condition given by Eq.~(\ref{eq:SO}) must be
fulfilled in each step of the iteration procedure.
We adopt a method of solving the coupled clusters equations through an unconstrained minimisation
of a variational functional which imposes the SO condition approximately by means of a penalty
term \cite{szalewicz82,szalewicz83a,szalewicz83b}. To this end, we employed the super-weak orthogonality (SWO) functional introduced in Ref.~\cite{wenzel86}. In the
case of the FCCD theory it has the following form 
\begin{equation}
\label{eq:SWO}
\begin{split}
\mathcal{J}_{\alpha\beta}^s[\tilde{\tau}]&=
  \langle\tilde{\tau}|f(1) + f(2) -\epsilon_\alpha - \epsilon_\beta |\tilde{\tau}\rangle \\
  &-2\operatorname{Re}\langle\tilde{\tau}|\bar R_{\alpha\beta}^s[{\bm \tau}^{[n-1]}]\rangle \\
  &+\Delta_1^{\alpha\beta}\langle\tilde{\tau}|p(1)+p(2)|\tilde{\tau}\rangle\\
  &+\Delta_2^{\alpha\beta}\langle\tilde{\tau}|p_e(1)+p_e(2)|\tilde{\tau}\rangle\\
  &+\Delta_3^{\alpha\beta}\langle\tilde{\tau}|p(1)p(2)|\tilde{\tau}\rangle,
\end{split}
\end{equation}
where the bar in $\bar R^s_{\alpha\beta}[{\bm \tau}^{[n-1]}]$ indicates that the one-electron SO projectors
appearing in the definition of $L^s_{\alpha\beta}$ in Eq.~(\ref{eq:rhs}) are omitted, and $p_e(1)$ is defined through Eq. (\ref{eq:pchi}) with orbital energies $\epsilon_\alpha$ multiplying each $\phi_\alpha(1)$ term.

No more than three-electron integrals are necessary within the SWO framework.
The minimisation of $\mathcal{J}_{\alpha\beta}^s[\tilde{\tau}]$ is performed only with
respect to the linear coefficients in the expansion of the trial function $\tilde{\tau}$ in terms of a set of fixed
basis functions. Therefore, finding a minimum of the functional is equivalent to solving a set of linear
equations. Additionally, after each step of the iteration procedure we perform an approximate projection of each pair
function \cite{szalewicz84b}. The strong-orthogonality projector is restricted to the space spanned by the geminal basis
set (SWO with projection technique, SWOP).

The last three terms in Eq.~(\ref{eq:SWO}) constitute a penalty function which increases
the value of the functional if any SO-violating components are present in the trial function.
We adopt the formulae from Ref.~\cite{wenzel86} for the parameters $\Delta_i^{\alpha\beta}$
\begin{equation}
\label{eq:SWO2}
\begin{split}
\Delta_1^{\alpha\beta}&=\epsilon_\alpha+\epsilon_\beta-\epsilon_\mathrm{HO}+\eta \\
\Delta_2^{\alpha\beta}&= -1 \\
\Delta_3^{\alpha\beta}&=2\epsilon_\mathrm{HO}- \epsilon_\alpha-\epsilon_\beta, \\
\end{split}
\end{equation}
where $\epsilon_\mathrm{HO}$ is the energy of the highest occupied reference orbital, and $\eta>0$ is a parameter which
allows us to control the strength of the SO forcing. The value of this parameter is irrelevant in the
limit of the complete basis set but influences the results in any finite basis set.

All pair functions are expanded in a common set of primitive functions of the form (\ref{eq:postac}). The proper
permutational symmetry of the singlet and triplet pair functions is ensured by applying the (anti-)symmetriser
$\hat{A}^s_{12}=1+(2-s)\hat{P}_{12}$, where $\hat{P}_{12}$ interchanges the electron coordinates.
The positive exponents $a_i$ and $a_j$, $i\leq j$, in Eq.~(\ref{eq:postac}) are all possible pairs
(including repetitions) created out of an $n_a$-element set of exponents. The powers $u$, $v$, $t$ are all
distinct non-negative integers subject to the condition
$u+v+t\leq\Omega$.
In the case when $a_i=a_j$ an additional constraint, $u\leq v$, is assumed. As a result, the set of primitive functions
is completely specified
by a set of $n_a$ exponents and a single number $\Omega$.
The total number of symmetric basis functions used to expand the singlet pairs can be calculated from the formula
\begin{equation}
\label{eq:numberK}
K(n_a,\Omega)=
  n_a\,\kappa_1(\Omega) + \frac{n_a\left(n_a-1\right)}2\,\kappa_2(\Omega),
\end{equation}
with $\kappa_1(\Omega)$ and $\kappa_2(\Omega)$ defined as
\begin{equation}
\begin{split}
\kappa_1(\Omega) & = \left\lfloor(\Omega+2)(\Omega+4)(2\,\Omega+3)/24\right\rfloor, \\
\kappa_2(\Omega) & = (\Omega+1)(\Omega+2)(\Omega+3)/6, \\
\end{split}
\end{equation}
where $\lfloor x\rfloor$ denotes the floor function of $x$. The number of antisymmetric basis functions used for the
expansion of the triplet pairs
is expressed by a formula similar to 
Eq.~(\ref{eq:numberK}) but with the term $n_a\,\kappa_1(\Omega)$ 
replaced by $n_a\,\kappa_1(\Omega-1)$. This results from the fact that the primitive functions in
Eq.~(\ref{eq:postac}) with $a_i=a_j$ and $u=v$ vanish after antisymmetrisation.

The SCF orbitals of the beryllium atom used in the FCCD calculations were calculated with the basis set in the form
\begin{align}
\label{hfbas}
 (4\pi)^{-1/2}\, r^s e^{-a_k r}.
\end{align}
where $s=0,\ldots,\omega$, and $k=1,2,\ldots,n_a$. This constitutes a set of approximations, denoted SCF$(n_a,\omega)$,
to the exact SCF energy. The optimal exponents were found by variational minimisation of the SCF energy for a fixed
$n_a$, $\omega$. Several representative examples of the calculated SCF energies are given in Table \ref{tab:SCF}. The
estimated limit ($-14.573\,023\,168\,316\,399\,582$) comes from calculations with $a_k=1/2,1,2,\ldots,10$, $\omega=7$, and we
believe it to be accurate to more than 20 significant digits. 

\begin{table}
\caption{\label{tab:SCF}
SCF energies of the beryllium atom calculated with SCF$(n_a,\omega)$ basis sets along with the optimised
exponents ($a_k$), and the corresponding errors ($\Delta E_\text{SCF}$) with respect to the reference
value (see text). All values are given in atomic units.}
\begin{ruledtabular}
\begin{tabular}{cd{1.4}d{1.3}d{2.3}d{2.2}c}
& 
\multicolumn{1}{c}{$a_1$} &
\multicolumn{1}{c}{$a_2$} &
\multicolumn{1}{c}{$a_3$} &
\multicolumn{1}{c}{$a_4$} &
\multicolumn{1}{c}{$\Delta E_\text{SCF}$} \\
\hline
SCF$(2,7)$ & 1.139  & 6.377 &        &       & $2\times10^{-10}$ \\
SCF$(3,3)$ & 0.9089 & 3.000 &  8.918 &       & $9\times10^{-10}$ \\
SCF$(3,5)$ & 0.9276 & 3.032 & 10.05  &       & $2\times10^{-14}$ \\
SCF$(3,7)$ & 0.9562 & 3.544 & 11.88  &       & $5\times10^{-17}$ \\
SCF$(4,5)$ & 0.8859 & 2.476 &  5.658 & 16.05 & $7\times10^{-18}$ \\
\end{tabular}
\end{ruledtabular}
\end{table}

The nonlinear parameters of the Hylleraas basis set (\ref{eq:postac}) were not optimised in subsequent explicitly correlated calculations. Instead, they are fixed as all possible combinations of nonlinear parameters from a given SCF$(n_a,\omega)$ wavefunction (subject to the conditions detailed earlier in the text).

In some terms of Eqs. (\ref{eq:etot})-(\ref{eq:SWO}) we encounter three-electron integrals of the following general
form
\begin{align}
 \int \frac{d\textbf{r}_1}{4\pi}\int \frac{d\textbf{r}_2}{4\pi}\int
\frac{d\textbf{r}_3}{4\pi}\;r_1^{n_1}\,r_1^{n_2}\,r_3^{n_3}\,r_{12}^{n_4}\,r_{13}^{n_5}\,r_{23}^{n_6}\;
e^{-a r_1-b r_2-c r_3}.
\end{align}
In the present work they were calculated with help of the method developed by Pachucki and collaborators \cite{pachucki04a} based on
a family of recursive relations. However, let us mention that some combinations of the powers $n_i$ are not required in
the FCCD computations. In fact, in all terms of Eqs. (\ref{eq:etot})-(\ref{eq:SWO}) at least one of $n_4$, $n_5$ or
$n_6$ is always either minus one or zero. This is advantageous as it eliminates a significant portion of the integrals
and reduces the size of the integral files. Calculations of the integral files were performed within the quad-double arithmetic precision (QD library \cite{qdlib}, approximately 64 significant digits) while the explicitly correlated computations were accomplished in the standard Fortran quadruple arithmetic precision (approximately 32 significant digits).

\subsection{Orbital calculations}

For the purposes of this paper we separate the total energy of an atom into several contributions
\begin{align}
\label{compot1}
 E_{\mbox{\scriptsize tot}} = E_{\mbox{\scriptsize SCF}} + E_{\mbox{\scriptsize FCCD}} + \delta_{\mbox{\scriptsize S}} +
\delta_{\mbox{\scriptsize NF}} + \delta_{\mbox{\scriptsize FCI}},
\end{align}
where $E_{\mbox{\scriptsize SCF}}$ is the reference Hartree-Fock energy, $E_{\mbox{\scriptsize FCCD}}$ is the
FCCD energy as described in the previous section, and
\begin{align}
 &\delta_{\mbox{\scriptsize S}} = E_{\mbox{\scriptsize CCSD}} - E_{\mbox{\scriptsize CCD}}, \\
 &\delta_{\mbox{\scriptsize NF}} = E_{\mbox{\scriptsize CCD}} - E_{\mbox{\scriptsize FCCD}},
\end{align}
where $E_{\mbox{\scriptsize CCD}}$ denotes the energy of the coupled cluster method with double excitations, and $E_{\mbox{\scriptsize CCSD}}$ - with single and double excitations \cite{bartlett07}.
Furthermore, $\delta_{\mbox{\scriptsize FCI}}$ denotes the remaining correlation energy due to triply and quadruply
excited configurations. The rearrangements in Eq. (\ref{compot1}) are formally exact and provide a convenient basis for
a composite method. In fact, the first two terms ($E_{\mbox{\scriptsize SCF}}$ and $E_{\mbox{\scriptsize FCCD}}$) are by
far dominating in Eq. (\ref{compot1}) and thus must be computed to very high absolute accuracy. The remaining terms are
orders of magnitude smaller and can be calculated with the standard methods based on the algebraic approximation.

The orbital calculations of $\delta_{\mbox{\scriptsize S}}$, $\delta_{\mbox{\scriptsize NF}}$, and
$\delta_{\mbox{\scriptsize FCI}}$ were performed in the basis set of the Slater-type orbitals (STOs) optimised
specifically for the purpose of this work. Overall, their composition and preparation is similar as in Refs. \cite{lesiuk14a,lesiuk14b,lesiuk15} but
involve functions with the highest angular momentum ranging from $L=2$ to $L=7$ (further details can be obtained from
the authors upon request).

The orbital coupled cluster calculations were performed with the \textsc{Gamess} program package \cite{gamess1,gamess2}. The CCD program
of Piecuch and collaborators \cite{piecuch02} was modified to exclude the non-factorisable CCD terms and thus make the orbital
calculations directly comparable with the explicitly correlated FCCD method described earlier. Full CI (FCI)
calculation were performed with newly developed general FCI program \textsc{Hector} \cite{przybytek14} written by one of us (M.P.).

\section{Numerical results}

\subsection{Explicitly correlated calculations}

The remaining problem in calculation of the FCCD energy is the choice of the strong-orthogonality forcing parameter
$\eta$, see Eq. (\ref{eq:SWO}) and (\ref{eq:SWO2}). In Table \ref{tab:eta} we show results
of FCCD calculations with a representative reference function SCF(3,7). 
The pair energies are given by Eq. (\ref{eq:epair}) with the following function in ket:
\begin{itemize}
 \item $\tau_{\alpha\beta}^s$ - no projection,
 \item $q_2\,\tau_{\alpha\beta}^s$ the exact strong-orthogonality projection,
 \item $q_B\tau_{\alpha\beta}^s$ - the approximate projection restricted to the given geminal basis, Refs. \cite{szalewicz84b}.
\end{itemize}
The deviations from the strong-orthogonality are measured with help of the following quantity
\begin{align}
\label{eq:sos}
 S = \max_{\alpha,\beta,s} \frac{\langle \tau_{\alpha\beta}^s | p(1) + p(2) | \tau_{\alpha\beta}^s\rangle}{\langle
\tau_{\alpha\beta}^s | \tau_{\alpha\beta}^s\rangle},
\end{align}
which is obviously zero when the exact operator $q_2$ is used.

\begin{table*}
\caption{\label{tab:eta}
Dependence of the calculated total FCCD correlation energy on the parameter $\eta$ for selected
values of $\Omega$. The SCF basis set is SCF$(3,7)$ and $S$ measures the deviation from the strong-orthogonality condition, see Eq. (\ref{eq:sos}). All values are given in m\hartree.
}
\begin{ruledtabular}
\begin{tabular}{lrrrrr}
$\eta$ &
\multicolumn{1}{c}{$q_2\,\tau_{\alpha\beta}^s$} &
\multicolumn{2}{c}{$\tau_{\alpha\beta}^s$} &
\multicolumn{2}{c}{$q_B\,\tau_{\alpha\beta}^s$} \\
 & \multicolumn{1}{c}{$E$} & \multicolumn{1}{c}{$E$} & \multicolumn{1}{c}{$\log_{10} S$} & \multicolumn{1}{c}{$E$} & 
 \multicolumn{1}{c}{$\log_{10}S$} \\
\hline\\[-2.2ex]
\multicolumn{6}{c}{$\Omega = 2$} \\
\hline\\[-2ex]
$10^{6}$ & $-$90.625\,379\,201 &   $-$90.625\,341\,531 &   $-$6.1 &  $-$90.625\,313\,732  & $-$6.1 \\
$10^{4}$ & $-$92.020\,366\,249 &   $-$92.021\,944\,336 &   $-$5.3 &  $-$92.018\,869\,225  & $-$5.3 \\
$10^{2}$ & $-$92.435\,575\,222 &   $-$92.603\,864\,518 &   $-$3.9 &  $-$92.432\,844\,678  & $-$5.1 \\
$10^{0}$ & $-$92.464\,729\,391 &   $-$94.168\,577\,256 &   $-$0.5 &  $-$92.463\,225\,833  & $-$5.1 \\
$10^{-2}$& $-$92.487\,865\,631 &   $-$91.743\,645\,183 &   $-$0.1 &  $-$92.486\,653\,218  & $-$4.9 \\
$0$      & $-$92.541\,519\,617 &  7256.282\,371\,921 &   +0.3 &  $-$92.541\,588\,807  & $-$4.9 \\
\hline\\[-2.2ex]
\multicolumn{6}{c}{$\Omega = 6$} \\
\hline\\[-2ex]
$10^{6}$ & $-$92.988\,766\,089 &   $-$92.988\,784\,177 &  $-$11.8 & $-$92.988\,766\,087 &   $-$12.9 \\
$10^{4}$ & $-$92.988\,766\,695 &   $-$92.990\,573\,333 &  $-$7.8  & $-$92.988\,766\,697 &   $-$11.9 \\
$10^{2}$ & $-$92.988\,766\,717 &   $-$93.151\,221\,033 &  $-$3.8  & $-$92.988\,766\,726 &   $-$11.9 \\
$10^{0}$ & $-$92.988\,766\,721 &   $-$94.133\,618\,742 &  $-$0.4  & $-$92.988\,766\,718 &   $-$11.9 \\
$10^{-2}$& $-$92.988\,766\,741 &   $-$91.012\,330\,964 &  $-$0.1  & $-$92.988\,766\,665 &   $-$11.8 \\
$0$      & $-$92.988\,766\,607 & 300619.748\,845\,206 &  0.3  & $-$92.988\,766\,961 &    $-$11.6 \\
\hline\\[-2.2ex]
\multicolumn{6}{c}{$\Omega = 10$} \\
\hline\\[-2ex]
$10^{6}$ & $-$92.988\,771\,476 & $-$92.988\,789\,564 &  $-$11.8 & $-$92.988\,771\,476 &  $-$17.6 \\
$10^{4}$ & $-$92.988\,771\,476 & $-$92.990\,578\,115 &  $-$7.8  & $-$92.988\,771\,476 &   $-$15.2 \\
$10^{2}$ & $-$92.988\,771\,476 & $-$93.151\,225\,803 &  $-$3.8  & $-$92.988\,771\,477 &   $-$14.0 \\
$10^{0}$ & $-$92.988\,771\,476 & $-$94.133\,624\,138 &  $-$0.4  & $-$92.988\,771\,477 &   $-$13.9 \\
$10^{-2}$& $-$92.988\,771\,476 & $-$91.012\,337\,221 &  $-$0.1  & $-$92.988\,771\,477 &   $-$13.9 \\
$0$      & $-$92.988\,771\,476 & 359.724\,711\,204 &  +0.3  & $-$92.988\,771\,477 &    $-$13.9 \\
\end{tabular}
\end{ruledtabular}
\end{table*}

From Table \ref{tab:eta} one can see that the approach without any projection yields useful results only when very large
$\eta$ is used in the iterative procedure. However, even under this condition the stability of the method is poor and
the results depend heavily on the adopted value of $\eta$. Therefore, this approach is not recommended even in large
basis sets.

On the other hand, the approximate and exact projections give very similar results with the difference diminishing with
increasing $\Omega$. Even more importantly, for larger $\Omega$ the results depend very weakly on the adopted $\eta$
and it is reasonable to set $\eta=0$. This confirms the earlier recommendations from Ref. \cite{szalewicz84a}.

In Table \ref{tab:conv} we present results of MP2 and FCCD calculations with several SCF basis sets and with systematic
increase of $\Omega$. This allows to investigate the convergence of the results towards the complete basis set
limit. In general, the convergence rate depends significantly on the value of $n_a$ in the reference SCF wavefunction. 
The number of $a_i$, $a_j$ pairs in the basis set (\ref{eq:postac}) which is used to expand the pair functions scales quadratically with $n_a$. This means that the flexibility of the trial wavefunction increases quickly with $n_a$ as illustrated in Table \ref{tab:conv}. With the SCF(2,7) reference wavefunction the results are not converged even with $\Omega$ as large as 15. If we employ $n_a=3$ the convergence of the MP2 energy to 1 p\hartree{} is achieved with $\Omega=15$ and with $n_a=4$ it is sufficient to use $\Omega=10$ in order to reach the same level. In the latter case the convergence rate is close to exponential, e.g. an increase of $\Omega$ by one unit allows to
recover one additional significant digit. Taking this into account we assume that the values obtained with the
SCF$(4,5)$ basis set and the largest $\Omega$ available are accurate to within all digits shown in Table
\ref{tab:conv}. This gives $-$76.358\,249\,287 m\hartree{} and $-$92.988\,771\,482 m\hartree{} as our best estimates
of the MP2 and FCCD total pair correlation energies in the beryllium atom. We believe that the error of both these
values is no larger than 1 p\hartree{} ($10^{-12}$ \hartree{}).

\begin{table*}
\caption{\label{tab:conv}
Convergence of the MP2 and FCCD correlation energies with $\Omega$ for different
SCF$(n_a,\omega)$ basis sets. Unless stated otherwise, the energies were obtained with $\eta=0$. All values are given in m\hartree.
}
\begin{ruledtabular}
\begin{tabular}{llllll}
$\Omega$ &
\multicolumn{1}{c}{SCF$(2,7)$} &
\multicolumn{1}{c}{SCF$(3,3)$} &
\multicolumn{1}{c}{SCF$(3,5)$} &
\multicolumn{1}{c}{SCF$(3,7)$} &
\multicolumn{1}{c}{SCF$(4,5)$} \\
\hline\\[-2.2ex]
\multicolumn{6}{c}{MP2} \\
\hline\\[-2ex]
\phantom{1}4 &  $-$76.312\,058\,331 &  $-$76.354\,429\,971 &  $-$76.353\,733\,011 &  $-$76.354\,112\,775 & 
$-$76.355\,826\,310 \\
\phantom{1}5 &  $-$76.353\,482\,469 &  $-$76.357\,871\,937 &  $-$76.357\,822\,897 &  $-$76.357\,873\,524 & 
$-$76.358\,023\,944 \\
\phantom{1}6 &  $-$76.357\,716\,463 &  $-$76.358\,208\,177 &  $-$76.358\,205\,564 &  $-$76.358\,209\,133 & 
$-$76.358\,229\,597 \\
\phantom{1}7 &  $-$76.358\,163\,297 &  $-$76.358\,244\,644 &  $-$76.358\,244\,549 &  $-$76.358\,244\,486 & 
$-$76.358\,247\,719 \\
\phantom{1}8 &  $-$76.358\,231\,147 &  $-$76.358\,248\,823 &  $-$76.358\,248\,708 &  $-$76.358\,248\,682 & 
$-$76.358\,249\,173 \\
\phantom{1}9 &  $-$76.358\,242\,659 &  $-$76.358\,249\,369 &  $-$76.358\,249\,184 &  $-$76.358\,249\,204 & 
$-$76.358\,249\,279 \\
          10 &  $-$76.358\,246\,439 &  $-$76.358\,249\,473 &  $-$76.358\,249\,255 &  $-$76.358\,249\,272 & 
$-$76.358\,249\,287 \\
          11 &  $-$76.358\,247\,892 &  $-$76.358\,249\,507 &  $-$76.358\,249\,272 &  $-$76.358\,249\,282 & 
$-$76.358\,249\,287 \\
          12 &  $-$76.358\,248\,568 &  $-$76.358\,249\,521 &  $-$76.358\,249\,279 &  $-$76.358\,249\,285 & 
$-$76.358\,249\,287 \\
          13 &  $-$76.358\,248\,915 &  $-$76.358\,249\,529 &  $-$76.358\,249\,283 &  $-$76.358\,249\,286 & 
$-$76.358\,249\,287 \\
          14 &  $-$76.358\,249\,104 &  $-$76.358\,249\,533 &  $-$76.358\,249\,285 &  $-$76.358\,249\,287 & 
$-$76.358\,249\,287 \\
          15 &  $-$76.358\,249\,212 &  $-$76.358\,249\,535 &  $-$76.358\,249\,286 &  $-$76.358\,249\,287 & 
$-$76.358\,249\,287 \\
\hline\\[-2.2ex]
\multicolumn{6}{c}{FCCD} \\
\hline\\[-2ex]
\phantom{1}4 &  $-$92.963\,174\,714 &  $-$92.987\,900\,387 &  $-$92.987\,651\,796 &  $-$92.987\,688\,596 & 
$-$92.988\,687\,929 \\
\phantom{1}5 &  $-$92.986\,550\,626 &  $-$92.988\,732\,196 &  $-$92.988\,705\,338 &  $-$92.988\,697\,147 & 
$-$92.988\,767\,531 \\
\phantom{1}6 &  $-$92.988\,565\,205 &  $-$92.988\,76\,8763 &  $-$92.988\,767\,304\footnotemark[1] &  $-$92.988\,766\,607
& $-$92.988\,771\,278 \\
\phantom{1}7 &  $-$92.988\,740\,066 &  $-$92.988\,771\,250 &  $-$92.988\,771\,050 &  $-$92.988\,771\,149 & 
$-$92.988\,771\,468 \\
\phantom{1}8 &  $-$92.988\,760\,589\footnotemark[1] &  $-$92.988\,771\,646 &  $-$92.988\,771\,371 &  $-$92.988\,771\,438
& $-$92.988\,771\,480 \\
\phantom{1}9 &  $-$92.988\,766\,686 &  $-$92.988\,771\,771 &  $-$92.988\,771\,435 &  $-$92.988\,771\,467 & 
$-$92.988\,771\,481 \\
          10 &  $-$92.988\,769\,260 &  $-$92.988\,771\,822 &  $-$92.988\,771\,460 &  $-$92.988\,771\,476 & 
$-$92.988\,771\,482\footnotemark[2] \\
\end{tabular}
\end{ruledtabular}
\footnotetext[1]{calculated with $\eta=10^{-2}$}
\footnotetext[2]{calculated with $\eta=10^6$}
\end{table*}

It is also important to consider the adequacy of the SCF reference function when accessing the accuracy of the final
results. In fact, the Hylleraas functional utilised in the present work is variational only with the exact reference
function. As illustrated in Table \ref{tab:conv} smaller SCF basis sets tend to give pair correlation energies which
are below the exact limit. This can lead to a spurious overestimation of the final results. To avoid this we follow a
general rule-of-thumb that the error in the SCF energy (which is much easier to control) must be at least by an order of magnitude smaller than the desired accuracy in the pair energies. For example, the SCF(3,3) energy is accurate to 0.9 n\hartree{} which causes the corresponding FCCD energy to overshoot by about 0.3 n\hartree{} below the estimated exact limit.

\begin{figure}
\includegraphics[width=\linewidth]{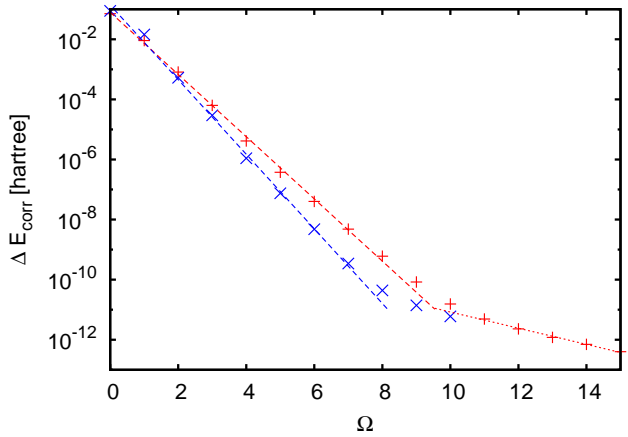}
\caption{\label{fig:conv} Convergence of the MP2 (red pluses) and FCCD (blue crosses) correlation energy with $\Omega$ for the
SCF$(3,7)$ basis set.}
\end{figure}

Finally, the convergence of the MP2 and FCCD correlation energies to the complete basis set limit is illustrated in
Fig. \ref{fig:conv}. One can see that the convergence rate of the FCCD energy is slightly faster than of MP2. Another
interesting phenomena is the pronounced change in the slope of the curve around $\Omega=8-10$. We do not have a
well-justified explanation of this behaviour but it is probably due to the fact that the same nonlinear parameters were
used in the SCF and Hylleraas pair functions (without re-optimisation). Other possible contributing factor is the
importance of three-particle cusp condition (at the coalescence point of two-electrons and the nucleus) which introduces
logarithmic singularities \cite{fock54,morgan86} in the exact pair functions.

The final results of our explicitly correlated calculations are summarised in Table \ref{tab:results}. The corresponding
results for helium atom and lithium cation/anion are also provided, together with data from Refs. \cite{patkowski07,bukowski99,przybytek09} which used to be the most accurate results available in the literature. The uncertainty of the present data ($\approx$ 1 p\hartree{}) constitutes an
improvement of roughly 5 orders of magnitude compared with previous works. The only exception is the lithium ion where
the straightforward iteration procedure converges only for small values of $\Omega$. For larger basis sets it becomes
oscillatory and finally diverges. This change in the behaviour usually occurred for $n_\mathrm{a}$ and $\Omega$ for which the
number of basis functions exceeded 400, and prevented us from generating more accurate results.

\begin{table}
\caption{\label{tab:results}
Correlation energies (in m\hartree{}) at different levels of theory for two- and four-electron atomic systems. The present results are shown in the first line while the best GTG results are collected in the second line.}
\begin{ruledtabular}
\begin{tabular}{llll}
& \multicolumn{1}{c}{MP2} & \multicolumn{1}{c}{FCCD} & \multicolumn{1}{c}{CCD\footnotemark[1]} \\
\hline\\[-2.2ex]
He     & $-$37.377\,474\,518\,9 & $-$42.017\,882\,917 \\
       & $-$37.377\,474\,52\footnotemark[2] & $-$42.017\,71\footnotemark[3] \\[1ex]
Li$^+$ & $-$40.216\,410\,043\,5 & $-$43.490\,592\,055 \\
       & $-$40.216\,32\footnotemark[3]      & $-$43.490\,46\footnotemark[3] \\[1ex]
Li$^-$ & $-$60.473\,978\,826\,7 & $-$71.293\,08       \\
       & $-$60.473\,971\footnotemark[4]     & $-$71.293\,022\footnotemark[5] &
$-$71.266\,072\footnotemark[5] \\[1ex]
Be     & $-$76.358\,249\,287\,3 & $-$92.988\,771\,482 \\
       & $-$76.358\,245\footnotemark[4]     & $-$92.988\,754\footnotemark[6] &
$-$92.961\,031\footnotemark[6] \\
\end{tabular}
\end{ruledtabular}
\footnotetext[1]{CCD and FCCD are equivalent for two-electron systems}
\footnotetext[2]{600-term GTG expansion, Ref.~\cite{patkowski07}}
\footnotetext[3]{150-term GTG expansion, Ref.~\cite{bukowski99}}
\footnotetext[4]{400-term GTG expansion (optimized for MP2), Ref.~\cite{przybytek09}.}
\footnotetext[5]{re-optimised 400-term GTG expansion (infinite-order functional), Ref.~\cite{przybytek09}.}
\footnotetext[6]{400-term GTG expansion (infinite-order functional), Ref.~\cite{przybytek09}.}
\end{table}

\subsection{Orbital calculations}

\begin{figure}
\includegraphics[width=\linewidth]{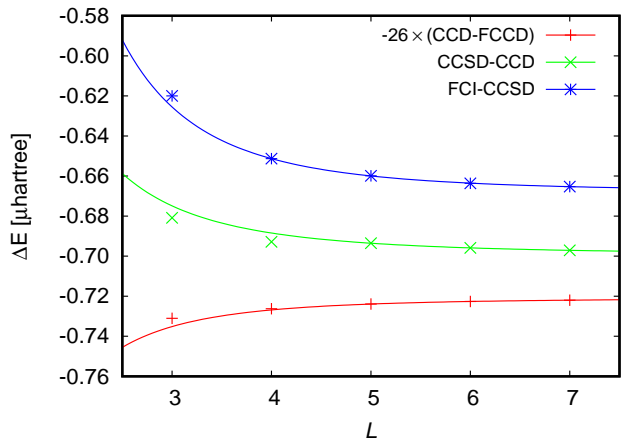}
\caption{\label{fig:convorb} Convergence of the $\delta_{\mbox{\scriptsize S}}$, $\delta_{\mbox{\scriptsize NF}}$, and
$\delta_{\mbox{\scriptsize FCI}}$ corrections to the complete basis set limit. The correction $\delta_{\mbox{\scriptsize
NF}}$ was multiplied by $-26$ to match the scale of the plot.}
\end{figure}

In Table \ref{tab:orbs} we present results of the calculations of the  $\delta_{\mbox{\scriptsize S}}$,
$\delta_{\mbox{\scriptsize NF}}$, and $\delta_{\mbox{\scriptsize FCI}}$ corrections using Slater-type orbitals
basis sets. The values from $L=3-7$ were extrapolated to the complete basis set limit with help of the following
three-point formula
\begin{align}
 A + \frac{B}{(L+1)^3} + \frac{C}{(L+1)^5}.
\end{align}
which was found to perform best for the FCCD pair energies (in comparison with the corresponding explicitly correlated
results). The quality of
the extrapolation is illustrated in Fig. \ref{fig:convorb}. One can see that the extrapolation formulae fit the results
from $L=4-7$ basis sets quite faithfully. The only exception is the basis set $L=4$ for $\delta_{\mbox{\scriptsize S}}$
which shows a considerable discrepancy making the extrapolated result less reliable. 

\begin{table}
\caption{\label{tab:orbs}
Corrections to the total correlation energy of the beryllium atom calculated within the STOs basis sets. The maximal
angular momentum in each basis set is provided in the first column. All values are given in m\hartree{}.}
\begin{ruledtabular}
\begin{tabular}{llll}
\multicolumn{1}{c}{$L$} & \multicolumn{1}{c}{$\delta_{\mbox{\scriptsize S}}$} &
\multicolumn{1}{c}{$\delta_{\mbox{\scriptsize NF}}$} & \multicolumn{1}{c}{$\delta_{\mbox{\scriptsize FCI}}$} \\
\hline\\[-2.2ex]
3 & $-$0.680\,857 & 0.028\,117 & $-$0.619\,981 \\
4 & $-$0.692\,823 & 0.027\,932 & $-$0.651\,325 \\
5 & $-$0.693\,542 & 0.027\,843 & $-$0.659\,907 \\
6 & $-$0.695\,871 & 0.027\,793 & $-$0.663\,636 \\
7 & $-$0.697\,089 & 0.027\,768 & $-$0.665\,259 \\
\hline\\[-2.2ex]
$\infty$ & $-$0.699\,299 & 0.027\,726 & $-$0.667\,195 \\
\end{tabular}
\end{ruledtabular}
\end{table}

The extrapolated values of all corrections are given in Table \ref{tab:orbs}. In Table \ref{tab:total} we provide a
short summary of the results of the present paper and give the final estimation of the total energy of the beryllium
atom. The errors of the respective components are estimated as twice the difference between the extrapolated results and the values in the largest basis set. The total error is about $5\,\mu$\hartree{} ($\approx1\,$cm$^{-1}$) compared with the result of Pachucki and
Komasa \cite{pachucki04b} which can be treated as a reference. This signifies that the present composite method is capable of reaching
the accuracy comparable to many spectroscopic measurements. Further in the text we discuss the error in each component
given in Table \ref{tab:total} and attempt to isolate the dominant source of the discrepancy. As argued in the previous
sections, the uncertainties in the SCF and FCCD energies are essentially negligible at this stage, as indicated in Table \ref{tab:total}.

\begin{table}
\caption{\label{tab:total}
Final prediction of the total energy of the beryllium atom. See the main text for details of the uncertainty estimation (shown in parentheses). The values without uncertainty estimation are accurate up to all digits quoted. The reference value is taken from Ref. \cite{pachucki04b}. All values are given in atomic units.}
\begin{ruledtabular}
\begin{tabular}{lr}
\multicolumn{1}{c}{contribution} & \multicolumn{1}{c}{value} \\
\hline\\[-2.2ex]
SCF  & $-$14.573\,023\phantom{(4)} \\
FCCD & $-$0.092\,989\phantom{(4)} \\
$\delta_{\mbox{\scriptsize NF}}$  & $+$0.000\,028\phantom{(4)} \\
$\delta_{\mbox{\scriptsize S}}$   & $-$0.000\,699(4) \\
total CCSD                        & $-$0.093\,660(4) \\[0.2ex]
$\delta_{\mbox{\scriptsize FCI}}$ & $-$0.000\,667(4) \\[0.2ex]
\hline\\[-2.2ex]
total energy & $-$14.667\,351(6) \\[0.2ex]
\hline\\[-2.2ex]
reference & $-$14.667\,356\phantom{(4)} \\
\end{tabular}
\end{ruledtabular}
\end{table}

The extrapolated value of the non-factorisable doubles correction ($\delta_{\mbox{\scriptsize NF}}$) agrees very well
with the result from Table \ref{tab:results} obtained independently with GTG expansions ($\delta_{\mbox{\scriptsize NF}}=0.027\,723$ m\hartree{}). 
The difference between these values is only about 3
n\hartree{} suggesting that both results are accurate to at least four significant digits. Moreover, as shown in Table
\ref{tab:orbs} the $\delta_{\mbox{\scriptsize NF}}$ correction stabilises quickly with increasing basis set size.
Therefore, we expect that in all practical applications it is sufficient to evaluate $\delta_{\mbox{\scriptsize NF}}$
with one-electron basis sets of a decent quality. In the present context, the uncertainty of $\delta_{\mbox{\scriptsize
NF}}$ does not contribute significantly to the overall error which is indicated in Table \ref{tab:total}.

Unfortunately, the same cannot be said about the singles correction, $\delta_{\mbox{\scriptsize S}}$. As mentioned
earlier, the convergence of $\delta_{\mbox{\scriptsize S}}$ towards the complete basis set limit is less regular than
for $\delta_{\mbox{\scriptsize NF}}$ or $\delta_{\mbox{\scriptsize FCI}}$ and thus the related extrapolation is not as
reliable. Therefore, we expect the extrapolated $\delta_{\mbox{\scriptsize S}}$ correction given in Table
\ref{tab:orbs} to be accurate only to two significant digits. In fact, the present result differs by as much as
$5\,\mu$\hartree{} from a more accurate value obtained in Ref. \cite{bukowski99} using an explicitly correlated variant of the CCSD theory. We believe that this discrepancy dominates the error in the total
energy of the beryllium atom given in Table \ref{tab:total}. To confirm this we replace $\delta_{\mbox{\scriptsize
S}}$ in Table \ref{tab:total} by the value from Ref. \cite{bukowski99} ($-0.705$ m\hartree{}). The total error then drops to
about $\approx\,0.1\,$cm$^{-1}$ which is an improvement by an order of magnitude. This shows clearly that the dominant
error to the total result given in \ref{tab:total} comes from inaccuracies in $\delta_{\mbox{\scriptsize S}}$.

Finally, the correction for the higher-order excitations ($\delta_{\mbox{\scriptsize FCI}}$) is of a similar magnitude
as $\delta_{\mbox{\scriptsize S}}$ but exhibits more regular convergence pattern towards the complete basis set limit.
While we do not have any reliable result in the literature to compare with directly, a comparison with
$\delta_{\mbox{\scriptsize NF}}$ allows us to claim that $\delta_{\mbox{\scriptsize FCI}}$ given in Table
\ref{tab:total} is accurate to three significant digits. In other words, the error in $\delta_{\mbox{\scriptsize FCI}}$
is of secondary concern in the present context. 

\section{Conclusions}

In this work we have reported the implementation and the first tests of a new composite method for accurate calculation
of energies of many-electron atoms. The dominant contribution to the energy has been calculated by using the explicitly
correlated factorisable coupled cluster theory. To expand the pair functions we have employed the Hylleraas basis set
and thus eliminated the need for optimisation of the nonlinear parameters at the correlated level. This allowed to compute pair
correlation energies of the beryllium atom with error smaller than 1 p\hartree{}, an improvement of several orders of
magnitude in comparison with the previous works. The remaining contributions to the total energy have been calculated
within the algebraic approximation employing large basis sets composed of Slater-type orbitals.

It is a natural and interesting question of how the present method can be used for heavier atoms retaining or improving
the current level of accuracy. In principle, the application of the theory to other many-electron atoms is
straightforward. However, the implementation is marred by difficulties related to proper treatment of angular factors
originating from $p$, $d$, $\ldots$ reference orbitals. Nonetheless, the Hylleraas basis set has been successfully
applied to (high $l$) excited states of the helium atom (see Ref. \cite{drake96} and references therein) and we believe that similar extensions are feasible here.

The present level of accuracy can be considerably improved if the correction due to single excitations
($\delta_{\mbox{\scriptsize S}}$) is computed with smaller uncertainty. First-quantised expressions for the explicitly
correlated CCSD model (where $\delta_{\mbox{\scriptsize S}}$ is included by construction) are well-known \cite{bukowski99}.
Unfortunately, their implementation requires four-integrals which are, in general, not available in
the Hylleraas basis set. Therefore, it is a considerable challenge to propose an approximate explicitly correlated CCSD
model where the most problematic four-electron integrals can be eliminated. This is similar to the idea of Bukowski et al. \cite{bukowski99} who proposed the factorisable quadratic CCSD model.

Another problem encountered for heavier atoms is calculation of energy contributions due to
higher-excitations from the reference determinant (pentuple, sextuple etc.) The most pragmatic approach is probably to
employ Quantum Monte Carlo FCI method \cite{booth09} which is capable of probing such large excitation
spaces stochastically. With the aforementioned improvements implemented we believe it would be possible to routinely
reach the accuracy of $0.1-0.01$ cm$^{-1}$ in calculation of the atomic energies. This also requires to include the
relativistic and quantum electrodynamics corrections, but as long as the atoms are not too heavy these effects can be
accounted for perturbatively. In this case the conventional calculations based on the algebraic approximation are
probably sufficient to deliver the desired accuracy.

\begin{acknowledgments}
We would like to thank B. Jeziorski for fruitful discussions, and for reading and commenting on the manuscript.
This research was supported by National Science Centre (NCN) Grant No. 2012/05/D/ST4/01271. 
\end{acknowledgments}


\begin{thebibliography}{000}
\small
\bibitem{roberts15} B. M. Roberts, V. A. Dzuba, and V. V. Flambaum, Annu. Rev. Nucl. Part. Sci. \textbf{65}, 63 (2015).
\bibitem{rosenband08} T. Rosenband, D. B. Hume, P. O. Schmidt, C. W. Chou, A. Brusch, L. Lorini, W. H. Oskay, R. E. Drullinger, T. M. Fortier, J. E. Stalnaker, S. A. Diddams, W. C. Swann, N. R. Newbury, W. M. Itano, D. J. Wineland, and J. C. Bergquist, Science \textbf{319}, 1808 (2008).
\bibitem{hunt14} N. Huntemann, B. Lipphardt, Chr. Tamm, V. Gerginov, S. Weyers, and E. Peik, Phys. Rev. Lett. \textbf{113}, 210802 (2014).
\bibitem{godun14} R. M. Godun, P. B. R. Nisbet-Jones, J. M. Jones, S. A. King, L. A. M. Johnson, H. S. Margolis, K. Szymaniec, S. N. Lea, K. Bongs, and P. Gill, Phys. Rev. Lett. \textbf{113}, 210801 (2014).
\bibitem{eikema97} K. S. E. Eikema, W. Ubachs, W. Vassen, and W. Hogervorst, Phys. Rev. A \textbf{55}, 1866 (1997).
\bibitem{korobov01} V. Korobov and A. Yelkhovsky, Phys. Rev. Lett. \textbf{87}, 193003 (2001).
\bibitem{pachucki06} K. Pachucki, Phys. Rev. A 74, 022512 (2006).
\bibitem{odom06} B. Odom, D. Hanneke, B. D'Urso, and G. Gabrielse, Phys. Rev. Lett. \textbf{97}, 030801 (2006).
\bibitem{gabrielse07} G. Gabrielse, D. Hanneke, T. Kinoshita, M. Nio, and B. Odom, Phys. Rev. Lett. \textbf{97}, 030802 (2006); Erratum: \emph{ibid}. 99, 039902 (2007).
\bibitem{hylleraas29a} E. A. Hylleraas, Z. Phys. \textbf{54}, 347 (1929).
\bibitem{hylleraas29b} E. A. Hylleraas, Z. Phys. \textbf{54}, 469 (1929).
\bibitem{drake94} G. W. F. Drake and Z.-C. Yan, Chem. Phys. Lett. \textbf{229}, 486 (1994).
\bibitem{puchalski06} M. Puchalski and K. Pachucki, Phys. Rev. A \textbf{73}, 022503 (2006).
\bibitem{szalewicz10} K. Szalewicz and B. Jeziorski, Mol. Phys. \textbf{108}, 3091 (2010).
\bibitem{mitroy13} J. Mitroy, S. Bubin, W. Horiuchi, Y. Suzuki, L. Adamowicz, W. Cencek, K. Szalewicz, J. Komasa, D. Blume, and K. Varga, Rev. Mod. Phys. \textbf{85}, 693 (2013).
\bibitem{sims71} J. S. Sims and S. A. Hagstrom, J. Chem. Phys. \textbf{55}, 4699 (1971).
\bibitem{sims07} J. S. Sims and S. A. Hagstrom, J. Phys. B \textbf{40}, 1575 (2007).
\bibitem{king93} F. W. King, J. Chem. Phys. \textbf{99}, 3622 (1993).
\bibitem{thakkar77} A. J. Thakkar and V. H. Smith, Jr., Phys. Rev. A \textbf{15}, 1 (1977).
\bibitem{frolov95} A. M. Frolov and V. H. Smith, Jr., J. Phys. B \textbf{28}, L449 (1995).
\bibitem{korobov02} V. I. Korobov, Phys. Rev. A \textbf{66}, 024501, (2002).
\bibitem{puchalski10} M. Puchalski and K. Pachucki, Phys. Rev. A \textbf{81}, 052505 (2010).
\bibitem{bartlett07} R. J. Bartlett and M. Musia\l, Rev. Mod. Phys. \textbf{79}, 291 (2007).
\bibitem{hill85} R. N. Hill, J. Chem. Phys. \textbf{83}, 1173, (1985).
\bibitem{kato57} T. Kato, Commun. Pure Appl. Math. \textbf{10}, 151, (1957).
\bibitem{byron66} F. W. Byron and C. J. Joachain, Phys. Rev. \textbf{146}, 1 (1966).
\bibitem{pan70} K. C. Pan and H. F. King, J. Chem. Phys. \textbf{53}, 4397 (1970).
\bibitem{pan72} K. C. Pan and H. F. King, J. Chem. Phys. \textbf{56}, 4667 (1972).
\bibitem{chalas77} G. Cha\l asi\'{n}ski, B. Jeziorski, J. Andzelm, and K. Szalewicz, Mol. Phys. \textbf{33}, 971 (1977).
\bibitem{szalewicz79} K. Szalewicz and B. Jeziorski, Mol. Phys. \textbf{38}, 191 (1979).
\bibitem{szalewicz82} K. Szalewicz, B. Jeziorski, H. J. Monkhorst, and J. G. Zabolitzky, Chem. Phys. Lett. \textbf{91}, 169 (1982).
\bibitem{szalewicz83a} K. Szalewicz, B. Jeziorski, H. J. Monkhorst, and J. G. Zabolitzky, J. Chem. Phys. \textbf{78},
1420 (1983).
\bibitem{szalewicz83b} K. Szalewicz, B. Jeziorski, H. J. Monkhorst, and J. G. Zabolitzky, J. Chem. Phys. \textbf{79},
5543 (1983).
\bibitem{szalewicz84a} B. Jeziorski, H. J. Monkhorst, K. Szalewicz, and J. G. Zabolitzky, J. Chem. Phys. \textbf{81}, 368
(1984).
\bibitem{szalewicz84b} K. Szalewicz, J. G. Zabolitzky, B. Jeziorski, and H. J. Monkhorst, J. Chem. Phys. \textbf{81}, 2723
(1984).
\bibitem{adamowicz77} L. Adamowicz and A. J. Sadlej, J. Chem. Phys. \textbf{67}, 4298 (1977).
\bibitem{adamowicz78a} L. Adamowicz and A. J. Sadlej, J. Chem. Phys. \textbf{69}, 3992 (1978).
\bibitem{adamowicz78b} L. Adamowicz, Int. J. Quantum Chem. \textbf{13}, 265 (1978).
\bibitem{hattig12} C. H\"attig, W. Klopper, A. K\"ohn, and D. P. Tew, Chem. Rev. \textbf{112}, 4 (2012).
\bibitem{kong12} L. Kong, F. A. Bischoff, and E. F. Valeev, Chem. Rev. \textbf{112}, 75 (2012).
\bibitem{tenno12} S. Ten-no, Theor. Chem. Acc. \textbf{131}, 1070 (2012).
\bibitem{fromm87} D. M. Fromm and R. N. Hill, Phys. Rev. A \textbf{36}, 1013 (1987).
\bibitem{remiddi91} E. Remiddi, Phys. Rev. A \textbf{44}, 5492 (1991).
\bibitem{harris97} F. E. Harris, Phys. Rev. A \textbf{55}, 1820 (1997).
\bibitem{pachucki04a} K. Pachucki, M. Puchalski, and E. Remiddi, Phys. Rev. A \textbf{70}, 032502 (2004).
\bibitem{Rychlewski:book} R. Bukowski, B. Jeziorski, and K. Szalewicz,
                   in {\em Explicitly correlated functions in chemistry 
                   and physics. Theory and applications}, edited by 
                   J. Rychlewski (Kluwer, Dordrecht, 2003), p. 185.
\bibitem{wenzel86} K. B. Wenzel, J. G. Zabolitzky, K. Szalewicz,  B. Jeziorski, and H. Monkhorst, J. Chem. Phys. \textbf{85}, 3964 (1986).
\bibitem{qdlib} D. H. Bailey, Y. Hida, and X. S. Li, \emph{"Libqd: quad-double / double-double computation package"}, http://crd-legacy.lbl.gov/~dhbailey/mpdist/ (2012), version 2.3.20.
\bibitem{lesiuk14a} M. Lesiuk and R. Moszynski, Phys. Rev. E \textbf{90}, 063318 (2014).
\bibitem{lesiuk14b} M. Lesiuk and R. Moszynski, Phys. Rev. E \textbf{90}, 063319 (2014).
\bibitem{lesiuk15} M. Lesiuk, M. Przybytek, M. Musial, B. Jeziorski, and R. Moszynski, Phys. Rev. A \textbf{91},
012510 (2015).
\bibitem{gamess1} M. W. Schmidt, K. K. Baldridge, J. A. Boatz, S. T. Elbert, M. S. Gordon, J. H. Jensen, S. Koseki, 
N. Matsunaga, K. A. Nguyen, S. Su, T. L. Windus, M. Dupuis, J. A. Montgomery, J. Comput. Chem. \textbf{14}, 
1347 (1993).
\bibitem{gamess2} \emph{"Advances in electronic structure theory: GAMESS a decade later"}, M. S. Gordon, M. W. Schmidt 
pp. 1167-1189, in \emph{"Theory and Applications of Computational Chemistry: the first forty years"}, C. E. Dykstra, G. 
Frenking, K. S. Kim, G. E. Scuseria (editors), Elsevier, Amsterdam, 2005.
\bibitem{piecuch02} P. Piecuch, S. A. Kucharski, K. Kowalski, M. Musia\l, Comm. Phys. Comm. \textbf{149}, 71 (2002).
\bibitem{przybytek14} M. Przybytek, FCI program \textsc{Hector}, 2014 (unpublished).
\bibitem{fock54} V. Fock, Izv. Akad. Nauk. SSSR, Ser. Fiz. \textbf{18}, 161 (1954).
\bibitem{morgan86} J. D. Morgan III, Theor. Chim. Acta \textbf{69}, 18 (1986).
\bibitem{patkowski07} K. Patkowski, W. Cencek, M. Jeziorska, B. Jeziorski, and K. Szalewicz, J. Phys. Chem. A \textbf{111}, 7611 (2007).
\bibitem{bukowski99} R. Bukowski, B. Jeziorski, and K. Szalewicz, J. Chem. Phys. \textbf{110}, 4165 (1999).
\bibitem{przybytek09} M. Przybytek, B. Jeziorski, and K. Szalewicz, Int. J. Quant. Chem. \textbf{109}, 2872 (2009).
\bibitem{pachucki04b} K. Pachucki and J. Komasa, Phys. Rev. Lett. \textbf{92}, 213001 (2004).
\bibitem{drake96} W. F. Drake: \emph{"High Precision Calculations for Helium"} in: Atomic, Molecular, and Optical Physics Handbook, ed. by G. W. F. Drake (AIP press, New York, 1996), pp. 154–171.
\bibitem{booth09} G. H. Booth, A. J. W. Thom, and A. Alavi, J. Chem. Phys. \textbf{131}, 054106 (2009).

\end{thebibliography}
\end{document}